# Uneven Use, Similar Trust? Generative AI in Denmark and Romania

*A descriptive two-country comparison of GenAI usage, trust in AI and institutional trust*

Iulian Lucau[1] Adelin George-Voicu[2] Simon Munk Christensen[3]

*GuidingAI, Sønderborg, Denmark[1]* *GuidingAI, Sønderborg, Denmark[2]* *GuidingAI, Sønderborg, Denmark[3]*

## Abstract

Generative AI (GenAI) has spread unevenly across Europe, and it is unclear whether higher use brings higher trust. This paper compares Denmark and Romania, the EU's highest and lowest GenAI users, and asks whether the difference in public and institutional trust between them corresponds to the difference in trust in AI. Drawing on trust, technology-acceptance and digital-divide theory, it is a descriptive, contrasting-case analysis of harmonised data: GenAI usage comes from Eurostat, trust in AI more generally from the KPMG/Melbourne global study and public and institutional trust from the Standard Eurobarometer. In 2025, 48.4% of Danes aged 16–74 had used GenAI in the past three months, against 17.8% of Romanians. Use falls with age and rises with education in both countries, and the relative gaps are far larger in Romania; the patterns remain visible within the published age-by-education categories. Trust in AI does not follow usage: Romanians trust AI as much as Danes or slightly more, although Denmark leads by 28 to 45 percentage points on trust in government, parliament and the legal system. The Denmark–Romania comparison thus does not support a simple country-level correspondence between general institutional trust and trust in AI; in KPMG's pooled model, trust in AI is tied most closely to confidence in AI governance. The observed usage difference is more consistent with disparities in general digital skills and opportunities for work- and education-related use than with differences in trust. However, the available data cannot identify the independent effects of these factors. The results rest on two cases and non-identical constructs, and call for comparable microdata and disclosure experiments.

***Keywords:*** generative AI, trust in AI, public and institutional trust, digital divide, technology acceptance, comparative analysis, Denmark, Romania

## 1. Introduction

Generative artificial intelligence has gone from a novelty to an everyday tool within a few years, yet its use is far from even across Europe. In 2025 Eurostat found that almost half of Danes (48.4%) aged 16–74 had used GenAI tools in the previous three months, against 17.8% of Romanians, the highest and lowest national shares in the EU (Eurostat 2025). But usage is only one part of the picture: whether people trust AI shapes their reliance on it, and the two do not necessarily align. The KPMG/Melbourne study, which asks both countries identical questions about AI in general, shows no higher trust in AI in Denmark despite its far higher GenAI usage (Gillespie et al. 2025, 109). To shape regulation and education effectively, it is essential to understand this difference.

The Danish and Romanian contrast is a useful one. Denmark has high public and institutional trust and strong self-rated digital skills, while Romania scores lower on both: in spring 2026, 67% of Danes against 25% of Romanians tended to trust their national parliament (Standard Eurobarometer 105, QA6), and 89% of Danes against 56% of Romanians consider themselves sufficiently skilled to use recent digital technologies in daily life (European Commission 2024, 22). This paper therefore asks: Does the difference in public and institutional trust between Denmark and Romania correspond to the difference in trust in AI?

Two sub-questions give the main question context. First, is the higher GenAI usage in Denmark accompanied by higher trust in AI, or does the link differ between the countries? Second, do age and education gaps in GenAI usage appear the same in both countries?

The paper draws on three harmonised sources: Eurostat for GenAI usage, the KPMG/Melbourne global study for trust in AI and related attitudes, and the Eurobarometer for public and institutional trust and positive attitude towards AI at work, supplemented by national survey data. Differences in survey design are a limitation (section 3.5), but they allow a comparison of two contexts that research has mostly examined separately.

### 1.1 Conceptual framework

Key concepts are defined below. Because the sources measure different things, the AI-related constructs from Trust in AI onwards are kept apart throughout the paper.

- Artificial Intelligence (AI): Computer systems designed to perform tasks requiring human-like cognitive abilities such as reasoning and pattern recognition.
- Generative AI (GenAI): AI models capable of 'creative' new synthetic content (text, images, code) by learning underlying data distributions based on immense datasets.
- AI Literacy: The technical knowledge and critical thinking required to understand, evaluate, and navigate AI technologies safely.
- Public and institutional trust: Citizens' confidence in state institutions, measured here by the share who tend to trust the national government, parliament and legal system (Standard Eurobarometer).
- Trust in AI: The willingness to be vulnerable to the outputs and actions of AI systems in general (Mayer, Davis, and Schoorman 1995, 712), measured by KPMG/Melbourne's trust scores (Gillespie et al. 2025). It is not specific to GenAI.
- Trust in GenAI: The same willingness directed at GenAI tools, such as trusting ChatGPT or AI-generated information. No harmonised measure exists, so it rests on national surveys (Reuters, PwC) and is context only.
- Acceptance of AI: Approval and embrace of the use of AI (KPMG/Melbourne's acceptance score).
- Positive attitude towards AI: A favourable overall view of AI or its effects, such as Eurobarometer's positive perception of AI in the workplace.
- Perceived benefits: The expectation that AI brings benefits (KPMG/Melbourne's benefits score and benefit-risk balance).
- Confidence in AI governance: Confidence that AI is developed, regulated and used responsibly, such as the perceived adequacy of laws and safeguards and confidence in government to develop and use AI in the public interest (KPMG/Melbourne).
- Opportunities for use: The settings (private, work, formal education) in which GenAI use occurs. Purpose-of-use data show where adoption occurs but are themselves forms of use, not independent measures of prior exposure (section 3.3).

## 2. Literature review

This review uses three mechanisms to turn the main question and sub-questions into expectations (section 2.4). The first concerns trust. Trust is the willingness to be vulnerable to the actions of another party, and it rests on perceived ability, benevolence and integrity (Mayer, Davis, and Schoorman 1995, 712, 717–19). Confidence in a parliament and trust in AI therefore draw on different judgments: the first on the integrity and benevolence of institutions, the second also on the ability of a system and the intentions of those who build and regulate it. KPMG's trust definition aligns with this model (Gillespie et al. 2025, 113), and its structural model makes the institutional pathway, confidence in AI governance, the strongest driver of trust in AI (B=.62), while the knowledge pathway of literacy, efficacy and training has a modest effect (B=.11; Gillespie et al. 2025, 59, 61).

The second mechanism concerns use. Technology acceptance research finds perceived usefulness more strongly linked to use than ease of use (Davis 1989, 319), so use can rise without a matching rise in trust. In KPMG's model, perceived benefits raise trust in AI and acceptance of AI (B=.23 and .22), while concern about risks has a smaller effect (B=−.08 for trust; Gillespie et al. 2025, 59, 61). Skills and opportunities for work- and education-related use enter this account as the ease and opportunity to use. The third mechanism concerns inequality. Digital-divide research treats access, skills and outcomes as successive levels of inequality and mostly examines sociodemographic and socioeconomic determinants (Scheerder, van Deursen, and van Dijk 2017, 1607). Age and education gaps are therefore where such divides should show, widest where access and skills are least evenly spread.

### 2.1 Adoption and socio-demographic divergence

GenAI adoption across Europe shows clear disparities in digital infrastructure, GenAI usage and socio-demographic dynamics. In Denmark the Digital Democracy Centre (DDC) reports rapid integration of GenAI in everyday life: between October 2023 and December 2025, Danes' private use rose from 33% to 62% and professional use from 24% to 56% (Sandberg et al. 2026, 7, 9). Reuters Institute data point the same way, although they cover another period and a different measure: Danish weekly use of the GenAI tools asked about rose from 15% to 34% between April 2024 and June–July 2025 (Simon et al. 2025, 17).

In Romania, the picture depends on the source. Eurostat's harmonised survey puts Romania last in the EU, with 17.8% of people aged 16–74 using GenAI tools in the previous three months, against 32.7% across the EU, and 44.1% among 16–24-year-olds, also the lowest share for that age group (Eurostat 2025). The KPMG/Melbourne study, which asked Danes and Romanians identical questions, finds a different ranking for AI use more broadly: 62% of Romanians and 57% of Danes use AI regularly or semi-regularly, that is, at least every few months (Gillespie et al. 2025, 21). Commercial surveys report still higher figures: AtlasIntel (2026) found that 83% of Romanians use AI in at least one context, and MKOR (2026, 5–8) reported overall AI familiarity of 70% among active adults, but sharply stratified by age (Gen Z 89%, baby boomers 39%).

Among Romanian students, 90.9% of a small sample reported prior AI tool experience, and 71% said ChatGPT significantly enhanced productivity (mean 4.15 out of five), while its perceived effect on critical thinking (2.91) and content originality (2.99) remained low (Constantin and Iacob 2025, 3362–8). Eurobarometer-based research (Sanda 2021) finds digital awareness gaps and older Romanians at higher risk of digital exclusion (Sanda 2021, 34, 38). More recent Eurobarometer data show the same gap in self-rated skills: 56% of Romanians consider themselves sufficiently skilled to use recent digital technologies, including AI, in daily life, against 89% of Danes, and younger and more educated respondents are the most likely to report sufficient skills (European Commission 2024, 9; 22).

### 2.2 Public and institutional trust, trust in AI and AI literacy

Public and institutional trust and trust in AI are not inherently related as "dual-trust": trust in institutions does not automatically carry over to trust in AI, and high public and institutional trust does not imply uncritical acceptance of synthetic AI material. Institutions do matter, however: in KPMG/Melbourne's pooled cross-national model, confidence in the entities developing AI and the perceived adequacy of AI regulation are the strongest drivers of trust in AI (B=.62), but these items measure confidence in AI governance rather than general public and institutional trust (Gillespie et al. 2025, 61).

Denmark illustrates this lack of "dual-trust". Danish society has high trust in public institutions and robust trust in news media, yet trust in AI is more moderate and conditional (Sandberg et al. 2026, 38). In the Reuters data, 72% of Danes trust the news in their country but only 32% trust ChatGPT (trust in GenAI), even though that is the second-highest share among the six countries surveyed (Simon et al. 2025, 27). In PwC's survey, 59% of Danes trust AI-generated information "to some extent" and 12% to a high degree (PwC 2026, 8), while the share with a positive overall attitude towards AI rose from 37% to 46% between 2024 and 2026 (PwC 2026, 4). DDC finds that although its AI literacy index rose to 0.56 by late 2025, its trust in AI index remained at 0.41, alongside a neutral-to-negative societal attitude among 49% (Sandberg et al. 2026, 8–16); the two Danish surveys differ in measures and fieldwork dates. The Danish ScI-AI project found that trust acts as a central mediator between technical understanding of AI and the epistemic evaluation of output reliability, so that understanding AI resulted in a more critical approach to it (Agergaard et al. 2026, 1–12). More literacy thus does not by itself increase trust.

In Romania, the harmonised KPMG/Melbourne data show slightly higher trust in AI than in Denmark: 4.1 against 3.8 on a seven-point scale, with Romanians also rating the safeguards around AI as more adequate (4.2 against 3.7), an indicator of confidence in AI governance (Gillespie et al. 2025, 109). Eurobarometer points the other way on positive attitude towards AI at work: 55% of Romanians perceive robots and AI in the workplace positively, against 86% of Danes, and since 2017 the share has fallen by 4 percentage points in Romania while rising by 4 in Denmark (European Commission 2024, 30–1). Across both contexts, trust is domain-contingent: educational institutions and healthcare providers attract higher trust when using AI, whereas commercial technology developers and government regulators face persistent public skepticism (Sandberg et al. 2026, 40; MKOR 2026, 28), and Danes are comfortable with AI-generated recommendations in job applications and customer service (40%) but rarely in law and public case handling (15%; PwC 2026, 9).

### 2.3 AI disclosure and the calibration of trust

Where use is already widespread, as in Denmark, the question shifts from whether people trust AI to whether their trust is calibrated, that is, matched to what AI can reliably do. Disclosure is one tool for informing it, such as the transparency obligations in Article 50 of the EU AI Act (Regulation (EU) 2024/1689). The experimental evidence points to a “transparency dilemma”: across 13 experiments, disclosing AI use lowered trust in the person disclosing it, which Schilke and Reimann (2025, 1, 12) attribute to reduced perceived legitimacy. Similar penalties appear for news tasks (Mattis, Kieslich, and de Vreese 2026, 1115, 1129) and credibility (Morosoli et al. 2025, 1748), although in advertising the effect ran both ways (Koning and Voorveld 2025, 240), and focus groups ask for fuller disclosure of substantive uses than a bare label (Cools et al. 2026, 16–17). All of this evidence comes from US, Dutch and UK samples (Schilke and Reimann 2025, 6; Mattis, Kieslich, and de Vreese 2026, 1122; Koning and Voorveld 2025, 245), and none tests Denmark or Romania.

### 2.4 Theoretical expectations

The three mechanisms above give an expectation for the main question and one for each sub-question. For the main question, because trust in a parliament and trust in AI rest on different judgments, public and institutional trust should be at best weakly related to trust in AI, given the lack of "dual-trust" in Denmark and the domain-contingent nature of trust, while confidence in AI governance should matter more (Gillespie et al. 2025, 61); if the Standard Eurobarometer confirms lower public and institutional trust in Romania, trust in AI there being at least as high as in Denmark would be consistent with this. For the first sub-question, because use follows perceived usefulness while trust rests on perceived ability, benevolence and integrity, usage and trust in AI should not move together uniformly: where AI literacy is higher, more knowledge and use may come with a more critical stance (Agergaard et al. 2026, 1–12), and literacy is a modest driver of trust in KPMG's model (B=.11; Gillespie et al. 2025, 59), so Denmark should pair high usage with moderate trust in AI, while Romania should pair lower measured usage with trust in AI at least as high, as is typical of KPMG's emerging economies (Gillespie et al. 2025, 20, 31). For the second sub-question, if divides follow uneven access and skills, age gaps should be steeper in Romania, where overall use is lower: among 16–24-year-olds Romania's rate is already 2.5 times its overall rate (44.1% against 17.8%), against 2.0 times for the EU (63.8% against 32.7%) (Eurostat 2025).

## 3. Methodology

### 3.1 Design

This paper is a comparative secondary analysis of two EU member states, Denmark and Romania. The two differ markedly in public and institutional trust (Standard Eurobarometer) and in self-rated general digital skills (89% and 56% consider themselves sufficiently skilled in daily life; European Commission 2024, 22), although not in AI knowledge (Gillespie et al. 2025, 109). Because usage differs sharply while trust in AI may not, the design is best described as a contrasting-case design; most-different-systems logic (Przeworski and Teune 1970) applies only to trust in AI as the outcome, where very different contexts may produce similar levels. The aim is to describe the patterns and test whether they are consistent with the literature, not to establish causal effects. As all data are public and anonymised, there is no ethical approval to consider.

### 3.2 Data and evidence hierarchy

No single survey measures usage, trust and skills in both countries, so sources are sorted into two tiers. Tier 1 is harmonised cross-national data with identical questions in both countries, and is the only basis for direct comparison: Eurostat's ICT survey (isoc_ai_iaiu), the source of all GenAI usage figures, by age and education in 2025; the KPMG/Melbourne global study (November 2024–January 2025), whose online panels of about 1,000 respondents per country are the source of trust in AI, acceptance of AI, perceived benefits and confidence in AI governance (Gillespie et al. 2025, 112); Special Eurobarometer 554 (Eurobarometer 101.4, ZA8844, April–May 2024; European Commission 2024, 1; TS3), whose AI module covers the positive attitude towards AI in the workplace and self-rated digital skills; the Standard Eurobarometer (waves 99, 101, 103 and 105, spring 2023 to spring 2026), whose QA6 items cover trust in the national government, parliament and legal system; and Eurostat's EU-SILC module on trust in others (ilc_pw03, a 0–10 rating used as a supplementary measure). Tier 2 is national and commercial data: the SDU DDC, the Reuters Institute and PwC/Epinion (Denmark), and MKOR and AtlasIntel (Romania). These add context within each country and are not used to compare the cases directly. Fieldwork runs from 2023 to 2026, so the main pairings are close in time: usage (Eurostat, 2025) is set against trust in AI (KPMG, November 2024 to January 2025), and public and institutional trust is checked across four Standard Eurobarometer waves that bracket KPMG's fieldwork.

### 3.3 Operationalisation

GenAI usage is the share of people aged 16–74 using GenAI in the past three months, taken from Eurostat, with purpose of use (private, work, formal education) as an indicator of the institutional contexts in which adoption occurs. Trust in AI more generally is taken from KPMG/Melbourne (not GenAI-specific): its trust in AI systems and trustworthiness scores (Gillespie et al. 2025, 109). Acceptance of AI, perceived benefits and confidence in AI governance (adequacy of laws and safeguards, confidence in government) come from the same study and are reported separately. Positive attitude towards AI is measured with Eurobarometer's positive perception of AI in the workplace. Trust in GenAI is only available as an approximate KPMG chart value and in national surveys (Reuters, PwC, DDC), which are not equivalent and serve as national context only. Digital skills are measured by a single self-rated Eurobarometer item on proficiency in using recent digital technologies, including AI, which captures general digital skills; AI literacy is measured separately by KPMG's AI knowledge, efficacy and training scores, because the two can point in different directions. Public and institutional trust is the share who tend to trust the national government, parliament and legal system (Standard Eurobarometer); trust in others, the mean rating in EU-SILC, is a supplementary measure outside this definition.

### 3.4 Analysis

For the age and education question, percentage-point gaps (16–24 vs. 55–74; high vs. low education within the 25–64 band) are calculated from Eurostat and compared across countries. To separate age from education, gaps are also compared within education groups and within age bands using Eurostat's cross-tabulated cells (Table 1); this is a stratified comparison, not a joint model. Because overall usage differs so much, relative gaps (ratios) are reported alongside them, and cells that Eurostat flags as unreliable are not interpreted. With only two countries, the main question and the usage-trust question are answered descriptively: public and institutional trust (Standard Eurobarometer) is set against trust in AI and the related KPMG and Eurobarometer constructs, and usage (Eurostat) against trust in AI, and consistency with the expectations in section 2.4 is assessed, with disagreement between indicators of the same factor (general digital skills against AI literacy) reported rather than resolved; no association between variables is tested. Differences smaller than the margin of error (about ±3 percentage points at $n \approx 1{,}000$; European Commission 2024, TS5) are not interpreted.

Levels of inference. All comparisons are between countries or between groups within countries, so they describe populations and cannot show whether the individuals who use GenAI more also trust it more or less (the ecological fallacy). The main question and the usage-trust question are therefore answered at the country level, and the conclusions concern Danes and Romanians as groups. Individual-level evidence is taken only from sources that analyse individual responses: KPMG's pooled structural model, which is not specific to either country (Gillespie et al. 2025, 59, 61), and Danish studies that relate experience, literacy and trust within the sample (Agergaard et al. 2026, 11–13). No such evidence for Romania is available in the sources reviewed.

### 3.5 Limitations

Romanian data are the weakest part, but not the only limited one. First, sources conflict: Eurostat reports 17.8% usage and AtlasIntel 83%, while KPMG reports 62% regular or semi-regular AI use for Romania against 57% for Denmark, reversing the Eurostat ranking (Gillespie et al. 2025, 21). The sources measure different things (GenAI in the last three months versus AI in any context at any frequency) and their samples differ: KPMG states that online panels in emerging economies over-represent the university-educated, urban and young, and 56% of its Romanian sample hold a degree, against 45% of its Danish sample (Gillespie et al. 2025, 107–8). Only Eurostat represents the whole population, so it is the benchmark for usage and KPMG for trust in AI; AtlasIntel and MKOR figures are context only, and as a sensitivity range Romanian usage is reported as 17.8%, 62% and 83%, with each construct labelled. Second, outside Eurostat, Eurobarometer and KPMG, Romanian figures come from commercial surveys with limited methodological transparency, although Danish sources have limits too: Reuters uses a non-probability online panel (Simon et al. 2025, 12), and PwC's survey has about 1,000 respondents (PwC 2026, 13). Third, Eurostat first asked the GenAI question in 2025, so no Tier 1 usage trend exists; the one available trend, Eurobarometer's positive attitude towards AI in the workplace since 2017, moved in opposite directions (Denmark +4, Romania −4 percentage points; European Commission 2024, 31). Fourth, Eurobarometer samples of roughly 1,000 respondents per country limit precision for age-by-education splits, and Denmark's video interviewing may add a mode difference (European Commission 2024, TS3; TS5). Fifth, sources were collected between 2023 and 2026 in a fast-moving field, so some differences may reflect timing: the main pairings are close in time and public and institutional trust is stable across four Standard Eurobarometer waves (section 4.2), but PwC's series shows Danish attitudes becoming more positive between 2024 and 2026 (37% to 46%; PwC 2026, 4), so KPMG's late-2024 figure for Denmark may understate its 2026 level. Sixth, the Danish EU-SILC series on trust in others has a break in 2018 and is not used to rank the countries, and Standard Eurobarometer trust is a yes-no item rather than a scale. Seventh, skills rest on one self-rated Eurobarometer item on general digital skills, which KPMG's AI knowledge and efficacy items do not corroborate, and prior exposure is not measured: purpose of use, occupation and place of residence show where adoption occurs but are themselves forms of use (section 3.3). Finally, self-reports are open to social desirability and varying interpretations of “AI”, country and group averages cannot show individual-level relationships (section 3.4), age and education are compared within cross-tabulated cells but not modelled jointly, and two cases restrict conclusions to description.

## 4. Findings

### 4.1 Usage

Overall levels. GenAI usage is taken from Eurostat. In 2025, 48.4% of Danes aged 16–74 had used GenAI tools in the previous three months, against 17.8% of Romanians and 32.7% across the EU (Eurostat 2025). Denmark is thus 30.6 percentage points above Romania and 15.7 above the EU average, while Romania's rate is 37% of Denmark's. Danish use is mainly private (34.8% of all Danes used GenAI privately, 27.2% for work and 17.9% in formal education), and Romanian use is even more private: 14.9% of all Romanians used GenAI privately, 5.2% for work and 3.4% in formal education (EU: 25.5%, 15.4% and 9.3%; Eurostat 2025). The Romanian work and education rates are thus only 19% of the Danish ones, against 43% for private use, so the Danish lead is largest in work and study settings.

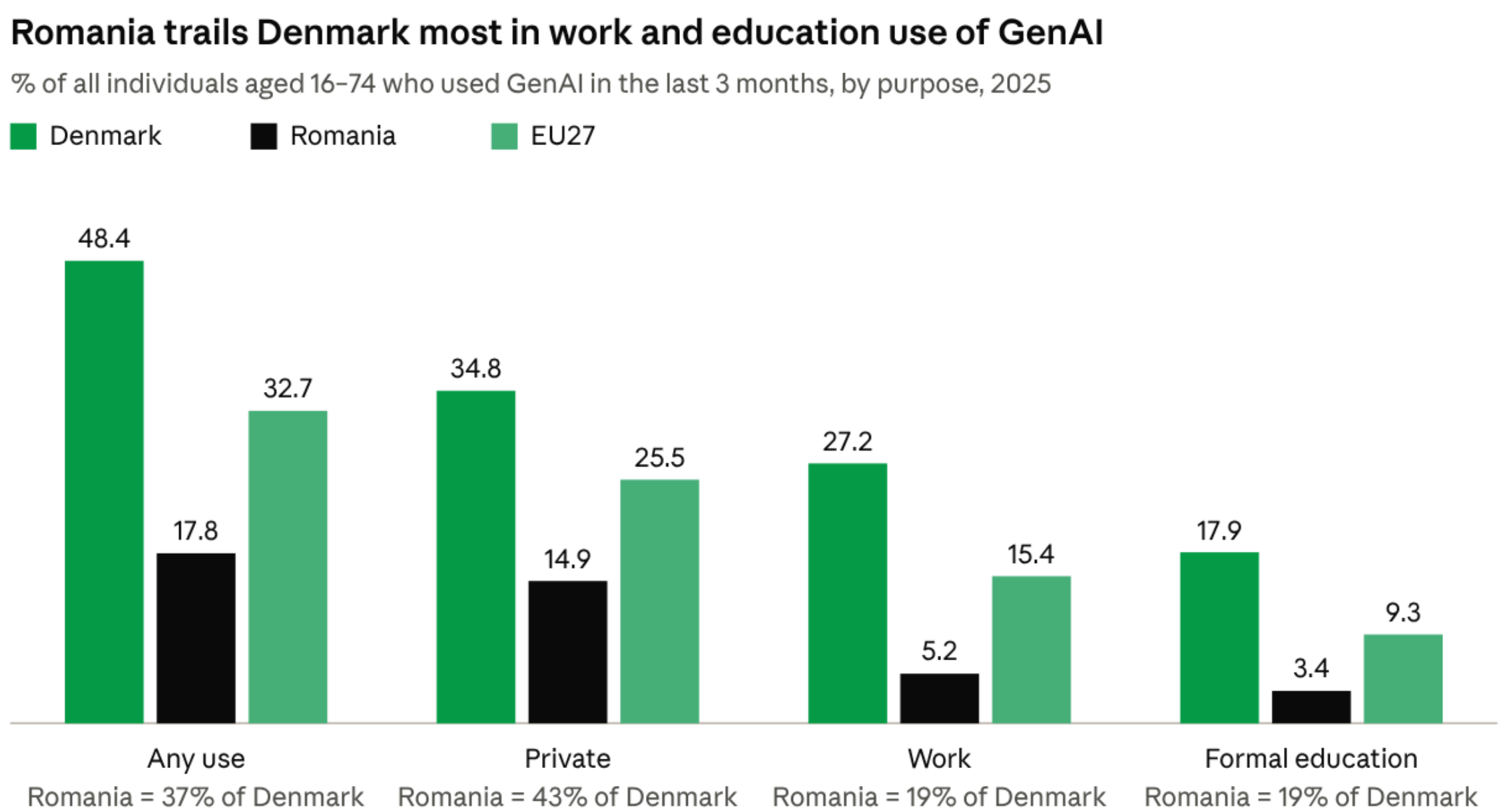

(Source: Eurostat (2025))

Age gaps. In both countries use falls steeply with age, but the shape differs. In Denmark, 77.6% of 16–24-year-olds used GenAI, falling to 31.6% among 55–64-year-olds and 14.8% among 65–74-year-olds; the gap between 16–24 and 55–74 (24.2%) is 53.5 percentage points, a ratio of 3.2. In Romania, 44.1% of 16–24-year-olds used GenAI, the lowest share in the EU for that group (63.8% EU-wide) but still 2.5 times Romania's overall rate, against 1.6 times in Denmark and 2.0 in the EU. Among 55–74-year-olds only 3.8% of Romanians used GenAI, against 24.2% of Danes and 12.0% across the EU, so the Romanian gap between 16–24 and 55–74 is 40.2 percentage points, smaller than Denmark's in absolute terms but a ratio of 11.5 against 3.2 in Denmark and 5.3 in the EU; Romanians aged 55–74 use GenAI at 22% of the national rate, against 50% in Denmark (Eurostat 2025). The commercial age stratification (section 2.1) points the same way, although it is not comparable.

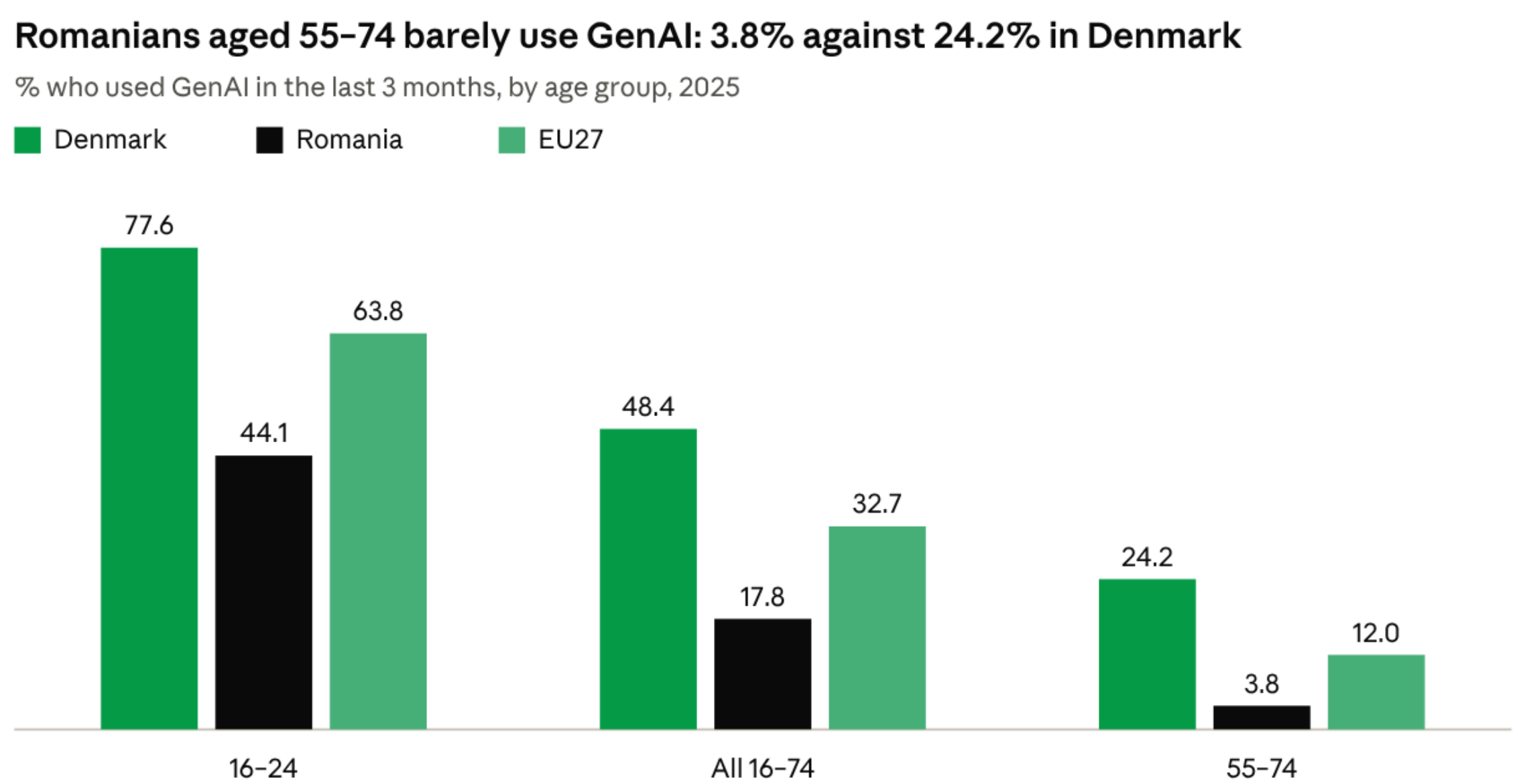

(Source: Eurostat (2025))

Education gaps. Within the 25–64 band, 26.6% of Danes with low formal education used GenAI, against 39.2% with medium and 63.5% with high education: a 36.9 percentage-point gap and a ratio of 2.4. Among 16–24-year-olds there is no education gradient in Denmark (76.7% low, 79.0% medium; the high-education cell is unreliable and not interpreted), so education matters there mainly after leaving education. In Romania the gradient is far steeper in relative terms: among 25–64-year-olds 2.6% with low, 11.9% with medium and 42.7% with high education used GenAI, a gap of 40.1 percentage points, slightly above Denmark's, but a ratio of 16.2 against 2.4 in Denmark and 3.9 in the EU. Among 16–24-year-olds a gradient exists in Romania (35.7% low, 48.8% medium, 57.6% high), unlike in Denmark, and among 55–74-year-olds 21.4% of the highly educated used GenAI against 0.1% of the low-educated, a cell so small that the ratio is not interpreted. Use by occupation follows a similar pattern: 80.4% of Danish ICT professionals and 79.5% of students used GenAI, against 56.4% of non-manual and 25.6% of manual workers, and in Romania 25.4% of ICT professionals, 26.4% of non-manual and 16.9% of manual workers; city residents use GenAI more than rural residents in both countries (60.5% against 36.8% in Denmark, 22.3% against 11.8% in Romania; Eurostat 2025). These breakdowns show where adoption occurs, not what explains it: work-related and education-related use are themselves forms of GenAI use, not independent measures of prior exposure.

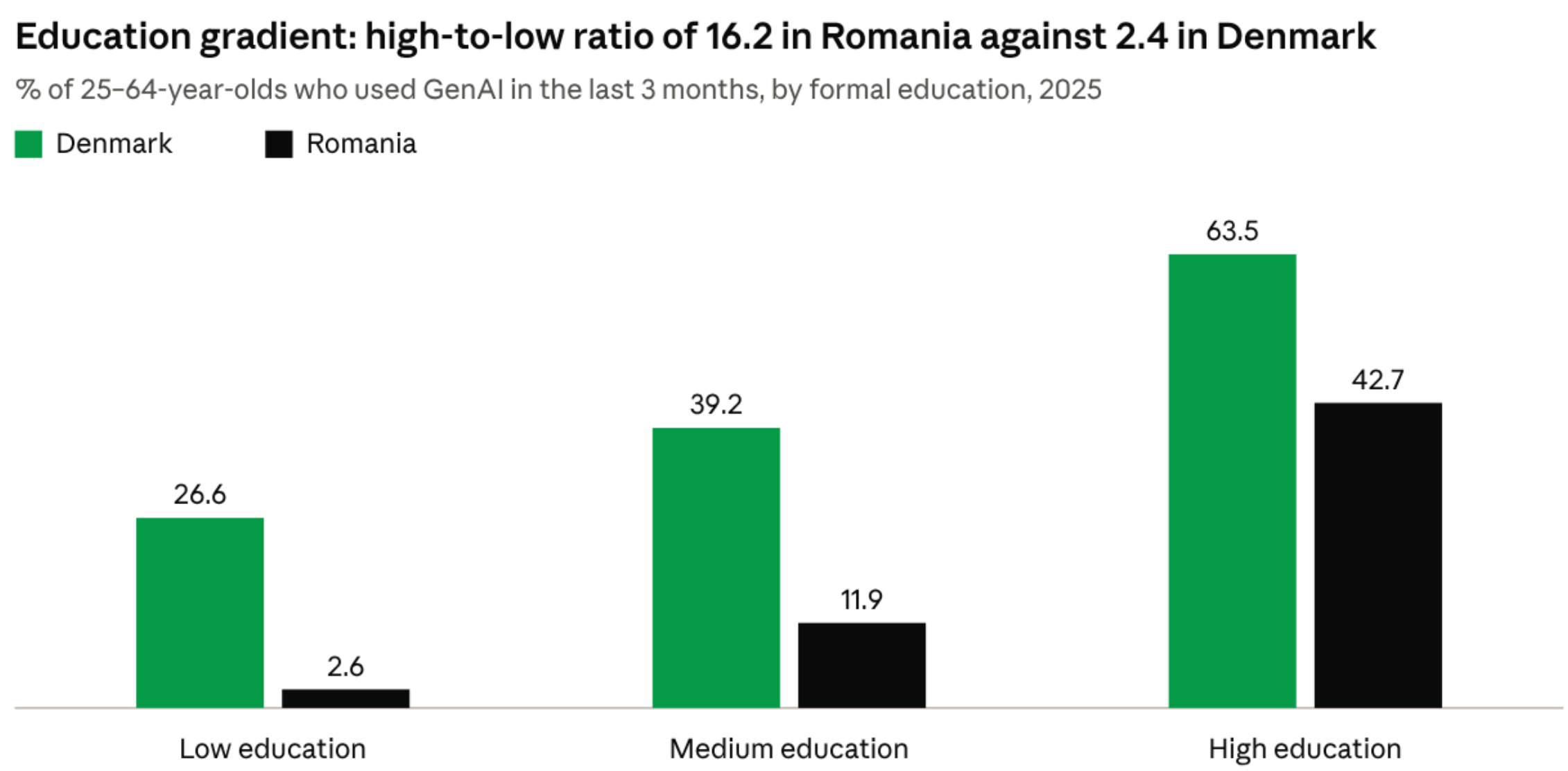

(Source: Eurostat (2025))

Age and education together. Eurostat also publishes use by age band and education jointly (Table 1), so the patterns can be checked within the published age-by-education categories. Romanian use is lower than Danish use in all nine cells, from 67% of the Danish rate among highly educated 25–54-year-olds to almost nothing among low-educated 55–74-year-olds (0.08%, a very small cell). The relative education gap is larger in Romania within every age band: the ratio of high to low use is 1.6 against 1.1 among 16–29-year-olds and 13.3 against 2.1 among 25–54-year-olds. The relative age gap between 25–54 and 55–74-year-olds is also larger in Romania at each education level (2.1 against 1.7 among the highly educated, 5.8 against 2.5 among those with medium education, and 42 against 2.1 among those with low education), although in percentage points the Romanian gaps are smaller (23 against 27 points by age among the highly educated) because levels are low. The steeper relative gradients in Romania therefore remain visible within the published age-by-education categories, although ratios near zero partly reflect floor effects and the groups are compared descriptively, not in a joint model.

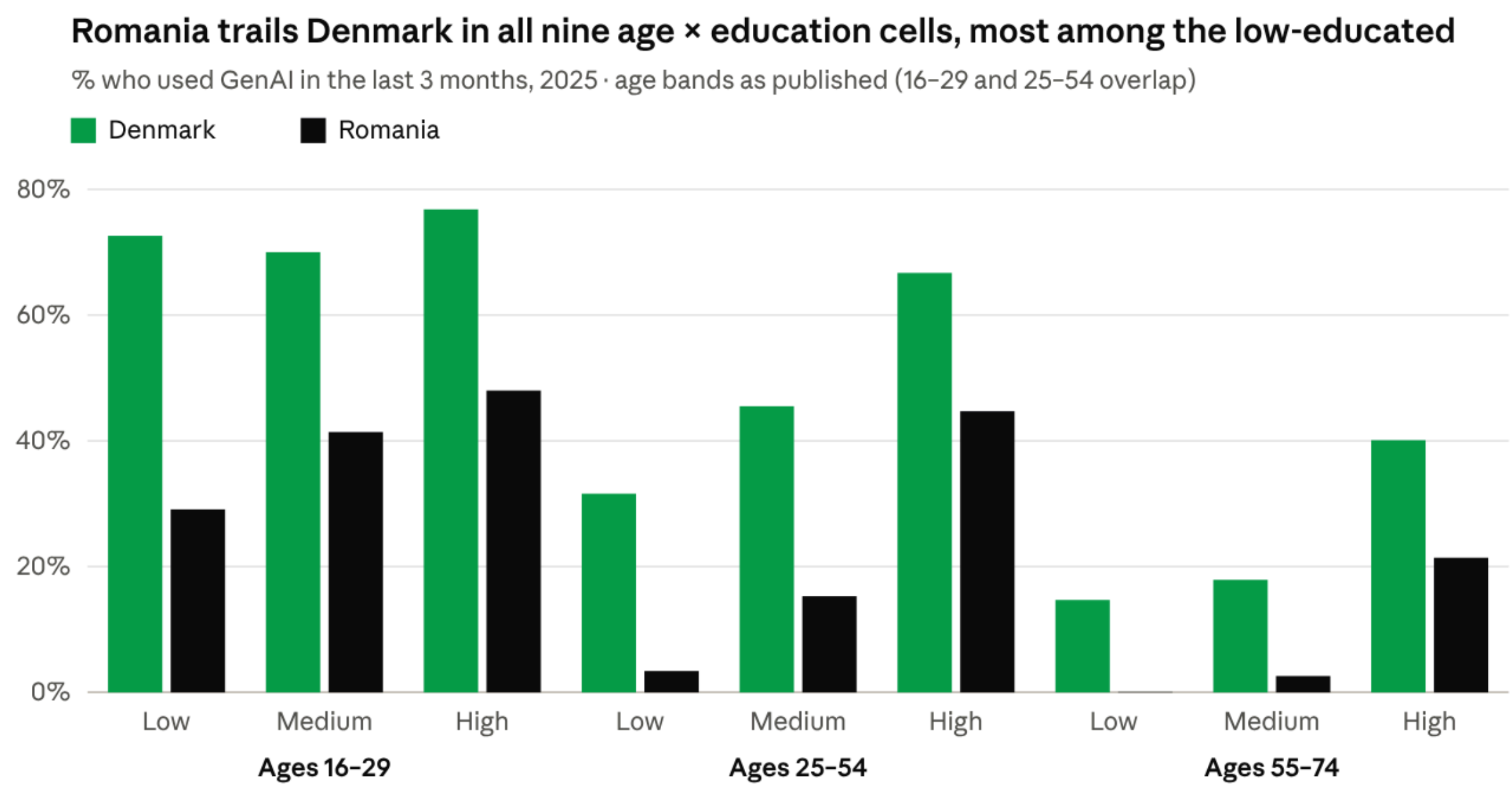

(Source: Eurostat (2025))

Table 1. GenAI use in the last three months by age band and education, 2025 (% of individuals)

| Age band | Education | Denmark | Romania | Romania as % of Denmark |
|---|---|---|---|---|
| 16–29 | Low | 72.6 | 29.1 | 40 |
| 16–29 | Medium | 70.0 | 41.4 | 59 |
| 16–29 | High | 76.8 | 48.0 | 62 |
| 25–54 | Low | 31.6 | 3.4 | 11 |

| Age band | Education | Denmark | Romania | Romania as % of Denmark |
|---|---|---|---|---|
| 25–54 | Medium | 45.5 | 15.3 | 34 |
| 25–54 | High | 66.7 | 44.7 | 67 |
| 55–74 | Low | 14.7 | 0.08 | 1 |
| 55–74 | Medium | 17.9 | 2.6 | 15 |
| 55–74 | High | 40.1 | 21.4 | 53 |

Source: Eurostat (2025), isoc_ai_iaiu. Age bands are as published and overlap (16–29 and 25–54 share ages 25–29). The Romanian cell for low-educated 55–74-year-olds is very small.

Summary for sub-question 2. In Denmark age is the dominant divide in usage (a 53.5-point gap), ahead of education (36.9 points) and place of residence; in Romania the age and education gaps are of similar absolute size (40.2 and 40.1 points). The ratio of youth use to overall use (2.5 against 1.6) and between the youngest and oldest groups (11.5 against 3.2) is higher in Romania, as is the relative education gap, which is about seven times larger (16.2 against 2.4 among 25–64-year-olds), consistent with the expectation in section 2.4 that relative gaps are larger where overall use is low. The patterns remain visible within the published age-by-education categories (Table 1): relative education gaps are larger in Romania in every age band and relative age gaps in every education group, although the groups are compared descriptively, not in a joint model.

### 4.2 Perceptions and trust

Trust in AI. Trust in AI more generally is taken from the KPMG/Melbourne study, which asked both countries the same questions. Romanians score slightly higher than Danes on trust in AI systems (4.1 against 3.8 on a seven-point scale) and on their trustworthiness (4.8 against 4.4), and the share willing to trust AI is 45% against 41% (Gillespie et al. 2025, 32; 109). Acceptance of AI is similar (high acceptance 36% against 33%), as is the benefit-risk balance (41% against 40% see benefits outweighing risks), although Romanians score higher on perceived benefits (3.7 against 3.3 out of five; Gillespie et al. 2025, 32; 45; 109). With about 1,000 respondents per country, percentage differences below about four points and score differences of 0.2 or less are not interpreted, so acceptance of AI and the benefit-risk balance are treated as equal. Trust in AI is thus similar in the two countries, with a small Romanian advantage that may partly reflect the larger share of graduates in the Romanian panel (56% against 45%; Gillespie et al. 2025, 107–8).

Confidence in AI governance. The clearest KPMG difference concerns governance. Similar shares want AI regulated (71% in Denmark, 70% in Romania; Gillespie et al. 2025, 49), but 47% of Romanians agree that current laws and safeguards are adequate, against 36% of Danes (Gillespie et al. 2025, 50). At the same time, half or more of Romanians lack confidence in their government to develop and use AI in the public interest (Gillespie et al. 2025, 55); read from the chart, confidence in the government is roughly 43% in Romania against 71% in Denmark (Gillespie et al. 2025, 57).

Skills and positive attitude towards AI at work. AI knowledge is identical (2.5 out of five), training is almost equal (34% against 33%) and Romanians report higher efficacy (4.7 against 4.1 out of seven) (Gillespie et al. 2025, 109), which does not match the Eurobarometer skills gap (89% of Danes against 56% of Romanians sufficiently skilled; European Commission 2024, 22). Eurobarometer also ranks the countries the other way on positive attitude towards AI at work: 86% of Danes perceive robots and AI in the workplace positively, against 55% of Romanians and 62% across the EU, and since 2017 the Danish share has risen by 4 points while the Romanian has fallen by 4 (European Commission 2024, 30–1). Differences in object, mode and sample mean that neither source gives the true level, so the finding depends on the construct.

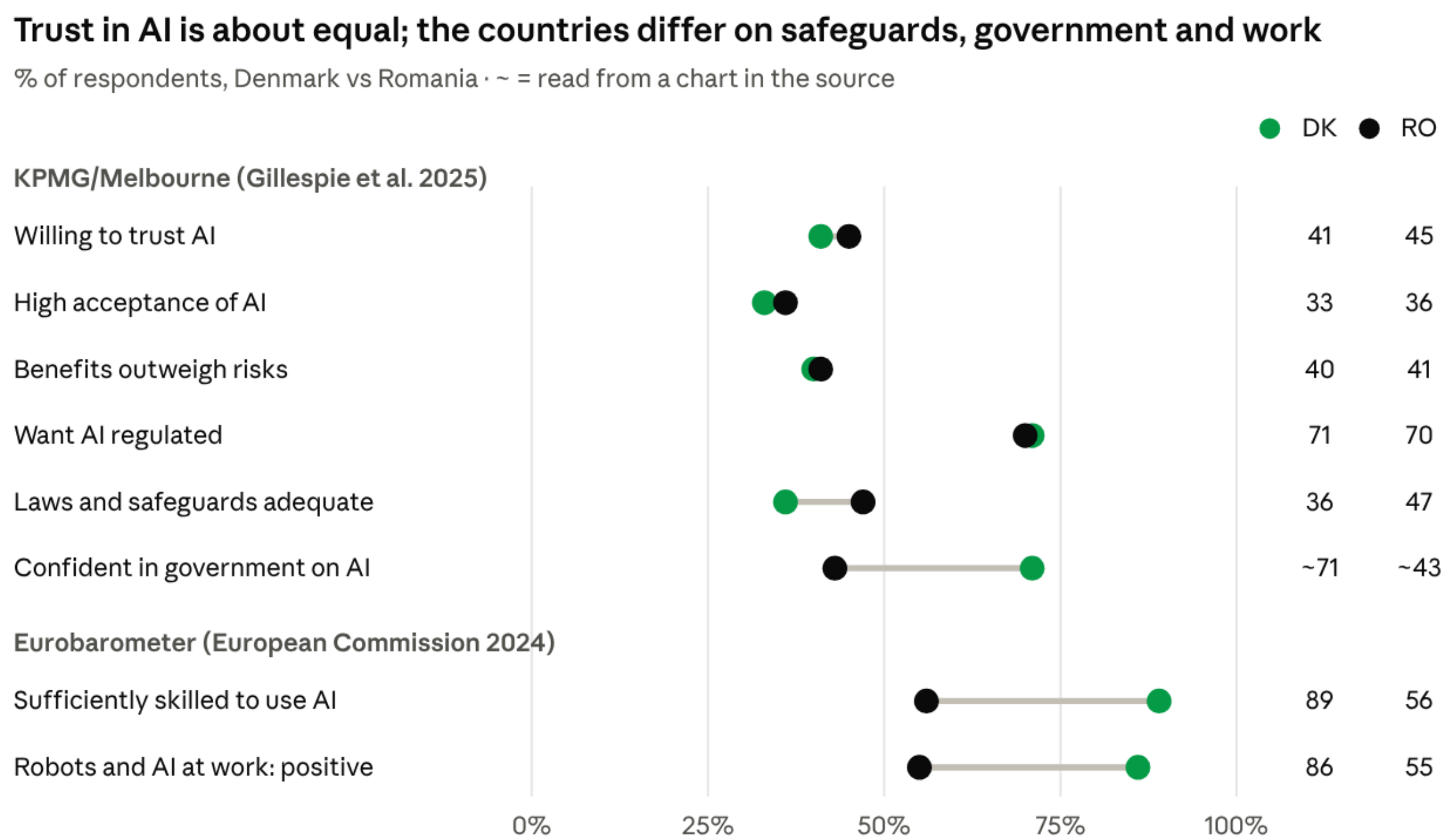

(Source: KPMG/Melbourne (Gillespie et al. 2025) and Special Eurobarometer 554)

Public and institutional trust. In four Standard Eurobarometer waves from spring 2023 to spring 2026, Denmark leads Romania widely and consistently. In spring 2026, 59% of Danes tended to trust their national government against 31% of Romanians, 67% against 25% for parliament, 90% against 45% for the legal system and 50% against 23% for political parties (EU27: 37%, 37%, 58% and 25%); over the four waves, Danish trust in government ranged from 54% to 65% and Romanian from 23% to 34%, a Danish lead of 25 to 33 points in every wave (Standard Eurobarometer 99–105, QA6). Eurostat's EU-SILC item on trust in others (0–10), which falls outside this definition, points the other way: 7.5 for Romanians and 5.7 for Danes in 2025 (EU 5.8; Eurostat ilc_pw03). However, the Danish series carries a break flag and falls from 8.3 in 2013 to 5.5 in 2018 while the Romanian rises from 6.4 to 7.0, so it is not used to rank the countries.

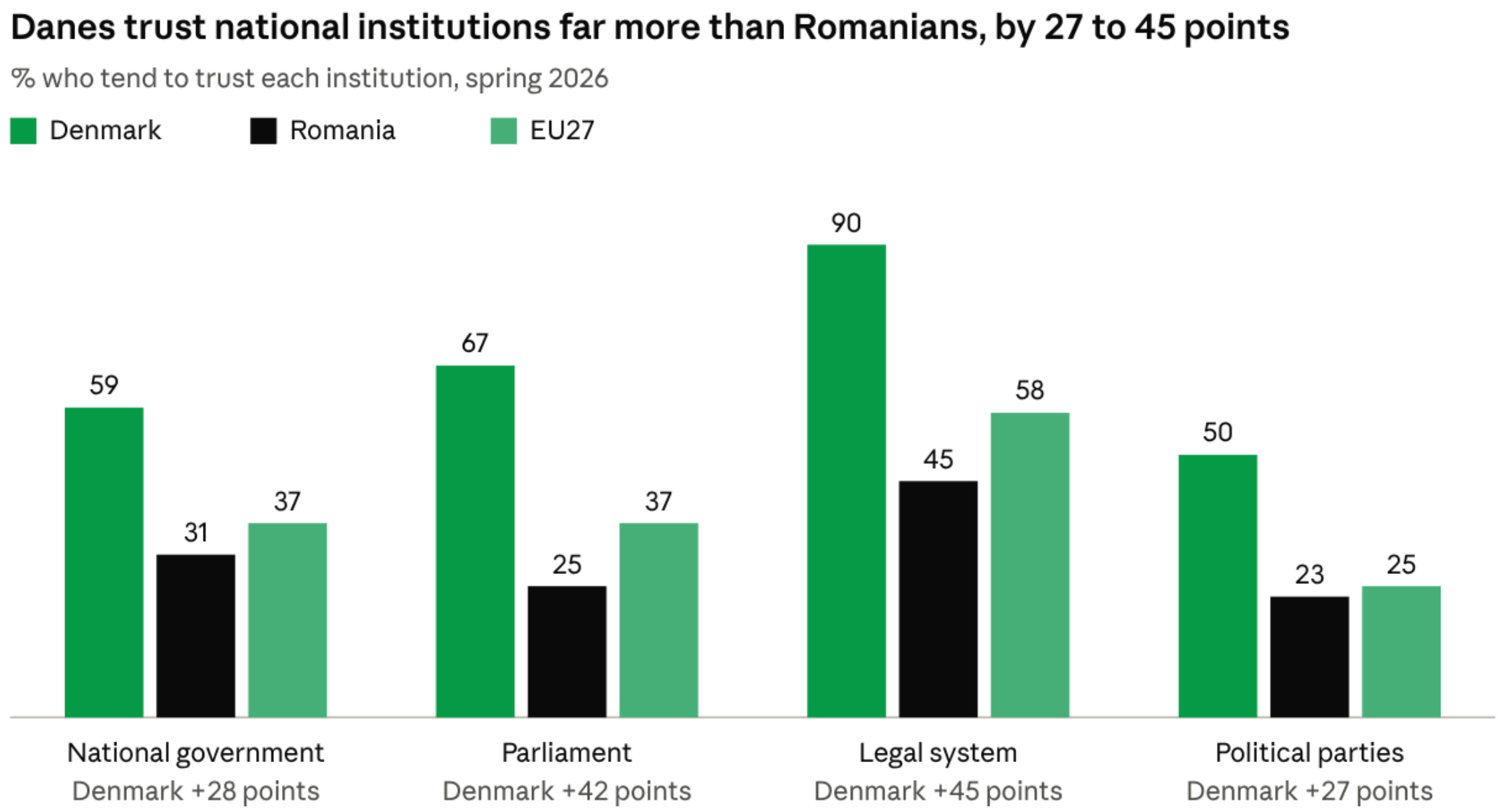

(Source: Standard Eurobarometer 105, QA6)

Sub-question 1. The link between usage and trust in AI differs by measure. Denmark has far higher Eurostat usage (48.4% against 17.8%) but no higher KPMG trust in AI, and on KPMG's broader measure Romanians report slightly higher regular use (62% against 57%; Gillespie et al. 2025, 21) and trust in AI. On the Eurobarometer item on positive attitude towards AI at work, however, higher Danish usage goes with a more positive attitude. With two countries and no microdata, this is a country-level comparison, and a within-country link cannot be tested.

Main question. The available evidence does not indicate clearly higher trust in AI in Denmark, despite Denmark's considerably higher GenAI usage and general institutional trust. Denmark has far higher trust in government, parliament and the legal system, yet Danes do not trust AI more than Romanians. Of the indicators of confidence in AI governance, confidence in the government to use AI mirrors public and institutional trust (higher in Denmark), whereas perceived adequacy of safeguards mirrors trust in AI (higher in Romania); KPMG's pooled model links confidence in AI governance to trust in AI (r=.54; B=.62; Gillespie et al. 2025, 55; 61). With two countries, this is a descriptive indication only.

Consistency with expectations. Both expectations in section 2.4 hold on the KPMG data: moderate Danish trust in AI despite high usage, and Romanian trust in AI at least as high despite much lower public and institutional trust. The Eurobarometer item on positive attitude towards AI at work is the exception.

## 5. Discussion

Usage and its drivers. The usage gap is large (48.4% against 17.8%; Eurostat 2025). The observed usage difference is more consistent with disparities in general digital skills and opportunities for work- and education-related use than with differences in trust. However, the available data cannot identify the independent effects of these factors. The Danish lead matches the Eurobarometer skills gap (89% against 56% sufficiently skilled; European Commission 2024, 9, 22), but this is a single self-rated item on general digital skills, and KPMG's AI knowledge (identical, 2.5 out of five) and efficacy items (Romanians 4.7 against 4.1 out of seven; Gillespie et al. 2025, 109) do not support it, so any Danish advantage may lie in general digital skills rather than AI literacy. Purpose-of-use data show where GenAI adoption occurs, but work-related and education-related usage are still forms of GenAI usage. They are not independent measures of prior exposure. Romanian use is also concentrated more sharply among the young and the educated (youngest-to-oldest ratio 11.5 against 3.2; education ratio 16.2 against 2.4), and these relative gaps remain larger within every age band and education group (Eurostat 2025), in line with national sources, though not at comparable levels (MKOR 2026, 11; Constantin and Iacob 2025, 3366; Sanda 2021, 34). This is consistency with, not proof of, a general-skills-and-opportunities explanation.

Trust in AI without a clear link to usage. The most surprising result is that the high-usage country does not trust AI more: on the harmonised KPMG measures trust in AI is similar or slightly higher in Romania (4.1 against 3.8; Gillespie et al. 2025, 109), and only the Eurobarometer item on positive attitude towards AI at work favours Denmark (86% against 55%; European Commission 2024, 30–31). Danish studies point the same way: the DDC trust in AI index stayed at a “low” 0.41 while private use rose from 33% to 62% (Sandberg et al. 2026, 7, 9), and Agergaard et al. (2026, 11–13) found experience with GenAI tools unconnected to trust, and trust in reliability lower among respondents who recognised its limits. This fits the expectation in section 2.4, although the larger graduate share in the Romanian panel (56% against 45%; Gillespie et al. 2025, 107–8) may raise its scores. This also bears on timing: Danish trust in AI stayed low while use rose, so the gap between KPMG's fieldwork and later usage data is unlikely to explain the pattern, and the individual-level link, which the country comparison cannot show, rests on these Danish studies and KPMG's pooled model (section 3.4).

Public and institutional trust and confidence in AI governance. In spring 2026 Denmark led Romania on trust in government (59% against 31%), parliament (67% against 25%) and the legal system (90% against 45%; European Commission 2023–2026, wave 105, 85, 87, 71), yet this does not carry over to trust in AI. Agergaard et al. (2026, 1, 12) describe Denmark as a high-trust context but find only moderate trust in GenAI, which supports the lack of "dual-trust" (section 2.2). What varies with trust in AI is confidence in AI governance: perceived adequacy of safeguards is higher in Romania (47% against 36%; Gillespie et al. 2025, 50), and confidence in AI developers and regulation is the strongest driver in KPMG's pooled model (B=.62; Gillespie et al. 2025, 61). Trust in AI thus seems to depend more on how AI itself is governed than on general institutional trust, although low Romanian confidence in government use of AI (Gillespie et al. 2025, 55, 57) still matters.

Implications. In Romania, access and digital-skills efforts should target the oldest and least educated, who are nearly absent from use (0.1% of low-educated 55–74-year-olds, a very small cell; Eurostat 2025). In Denmark, where use is widespread but trust in AI moderate, the issue is calibration. Danes want AI disclosed (87%; PwC 2026, 12) but rarely see labels (12% daily; Simon, Nielsen, and Fletcher 2025, 54), and the experiments in section 2.3 suggest that a bare label can lower trust while audiences want more than a label (Schilke and Reimann 2025, 12; Cools et al. 2026, 16–17). Article 50 of the AI Act may therefore work best with contextual disclosure, but this evidence is untested in Denmark or Romania.

Limits and further research. The comparison rests on two cases and constructs that differ between sources (section 3.5), so national figures are context only. Further research should use microdata with identical usage, trust and skills items (including AI-specific literacy and direct measures of prior exposure) in both countries to test the usage-trust link within countries and separate age from education. If Eurobarometer 101.4 microdata are accessed, a logistic regression per country of a positive attitude towards AI in the workplace on age, education and self-rated digital skills, using the supplied survey weights, would be a first step; no such analysis was conducted here. Disclosure formats should also be tested on Danish and Romanian audiences.

## 6. Conclusion

This paper asked whether the difference in public and institutional trust between Denmark and Romania corresponds to the difference in trust in AI. The Denmark–Romania comparison does not support a simple country-level correspondence between general institutional trust and trust in AI. In spring 2026 Denmark led Romania by 28 to 45 percentage points on trust in government, parliament and the legal system (Standard Eurobarometer 105, QA6), yet on the harmonised KPMG/Melbourne measures Romanians trust AI as much as Danes or slightly more (4.1 against 3.8 on a seven-point scale; Gillespie et al. 2025, 109), and only the Eurobarometer item on positive attitude towards AI at work favours Denmark. What varied with trust in AI was confidence in AI governance, such as the perceived adequacy of safeguards.

Two sub-questions give this result context. First, across the two countries higher usage did not go with higher trust in AI: Denmark combines widespread use with moderate and conditional trust in AI, and individual-level Danish studies suggest that experience and literacy do not by themselves raise it. The usage gap (48.4% against 17.8% in 2025; Eurostat 2025) is more consistent with disparities in general digital skills and opportunities for work- and education-related use than with differences in trust, but the available data cannot identify the independent effects of these factors. Second, age and education gaps exist in both countries but are far larger in relative terms in Romania, and the patterns remain visible within the published age-by-education categories (youngest-to-oldest ratio 11.5 against 3.2; education ratio 16.2 against 2.4). This is consistent with GenAI having spread mainly within a young and educated minority.

The findings suggest two challenges. In Romania, the task is to widen access and digital skills beyond the young and the highly educated. In Denmark, where use is widespread, it is to support well-calibrated trust, including disclosure that tells people more than a label, since experimental evidence from other countries shows that simple disclosure can lower trust (Schilke and Reimann 2025, 12). These conclusions are descriptive. They rest on two cases, on sources that measure different constructs, on weak and partly conflicting skills measures, on the lack of an independent measure of prior exposure and on aggregate data without microdata, so they show consistency with, not proof of, a general-skills-and-opportunities explanation. The next step is comparable microdata on GenAI use, trust in AI, AI literacy and prior exposure in both countries, and disclosure experiments on Danish and Romanian audiences.

### 6.1. A.I. statement:

AI, specifically Claude.ai Sonnet 5.5 (Anthropic, October 2026), was utilized in preparation, data sorting, structuring and reflections throughout paper creation. Visualization of data in sections 4.1-2 were created with Claude.ai Sonnet 5.5. Proof rereading, polishing and reviewing done manually. Topic, sources, analysis direction, review of output and content quality management were done by the authors for whom accountability lies for any errors.

## 7. References

### Academic literature

**Agergaard, Torben E., Kristian H. Nielsen, Rodrigo Labouriau, and Antoinette Fage-Butler. 2026.** "Exploring Trust and Literacy in Engagement With Generative AI and Science Information Behavior." *Media and Communication* 14: article 11357. https://doi.org/10.17645/mac.11357.

**Constantin, Alexandra, and Silvia-Elena Iacob. 2025.** "Higher Education in the Age of AI: ChatGPT's Impact on Students' Academic Activities." In *Proceedings of the 19th International Conference on Business Excellence 2025*, 3360–76. Sciendo. https://doi.org/10.2478/picbe-2025-0256.

**Cools, Hannes, Sophie Morosoli, Laurens Naudts, Karthikeya Venkatraj, Claes de Vreese, and Natali Helberger. 2026.** "'Transparency Is More Than a Label': Audiences' Information Needs for AI Use Disclosures in News." *Digital Journalism*, published ahead of print, August 1, 2026. https://doi.org/10.1080/21670811.2026.2700592. Page numbers refer to the article as published online (pp. 1–21).

**Davis, Fred D. 1989.** "Perceived Usefulness, Perceived Ease of Use, and User Acceptance of Information Technology." *MIS Quarterly* 13 (3): 319–40. https://doi.org/10.2307/249008.

**Koning, Britta, and Hilde A. M. Voorveld. 2025.** "Disclaimer! This Content Is AI-Generated: How AI-Disclosures Influence Trust in Advertisements and Organizations." *Journal of Interactive Advertising* 25 (3): 240–53. https://doi.org/10.1080/15252019.2025.2554149.

**Mattis, Nicolas, Kimon Kieslich, and Claes Holger de Vreese. 2026.** "Feeling Iffy About Generative AI: Investigating Audiences' Trustworthiness Perceptions of Task-Specific AI Disclosures." *Digital Journalism* 14 (7): 1115–35. https://doi.org/10.1080/21670811.2026.2703599.

**Mayer, Roger C., James H. Davis, and F. David Schoorman. 1995.** "An Integrative Model of Organizational Trust." *Academy of Management Review* 20 (3): 709–34. https://doi.org/10.5465/amr.1995.9508080335.

**Morosoli, Sophie, Emma van der Goot, Valeria Resendez, Claes de Vreese, and Natali Helberger. 2025.** "The Transparency Dilemma: An Experiment on How AI Disclosures Affect Credibility Perceptions and Engagement Across Topics." In *Proceedings of the Eighth AAAI/ACM Conference on AI, Ethics, and Society (AIES 2025)*, 1748–57. AAAI Press.

**Przeworski, Adam, and Henry Teune. 1970.** *The Logic of Comparative Social Inquiry*. New York: Wiley-Interscience.

**Sanda, Emanuel. 2021.** "Artificial Intelligence Applications and Impact on Consumers – With a View on Romanian Consumers' Perceptions and Attitudes Related to AI." *Studia Universitatis Babeș-Bolyai Negotia* 66 (2): 27–39. https://doi.org/10.24193/subbnegotia.2021.2.02.

**Scheerder, Anique, Alexander van Deursen, and Jan van Dijk. 2017.** "Determinants of Internet Skills, Use and Outcomes: A Systematic Review of the Second- and Third-Level Digital Divide." *Telematics and Informatics* 34 (8): 1607–24. https://doi.org/10.1016/j.tele.2017.07.007.

**Schilke, Oliver, and Martin Reimann. 2025.** "The Transparency Dilemma: How AI Disclosure Erodes Trust." *Organizational Behavior and Human Decision Processes* 188: article 104405. https://doi.org/10.1016/j.obhdp.2025.104405.

**Databases, datasets and reports**

**AtlasIntel. 2026.** ***The Digitalization of Public Services in Romania III*****. Survey of 2,002 adults, fieldwork June 17–21, 2026**, commissioned by Edge Institute. Results published in Edge Institute, “Artificial Intelligence and Work: The Private Sector, the Public Sector, and Those Without a Job,” July 29, 2026. https://edgeinstitute.ro/en/p/artificial-intelligence-and-work/.

**European Commission. 2023–2026.** *Standard Eurobarometer 99, 101, 103 and 105: Public Opinion in the European Union, Data Annexes, Spring 2023–Spring 2026*. Brussels: European Commission. Page numbers cited refer to the wave 105 annex.

**— . 2024.** *Special Eurobarometer 554: Artificial Intelligence and the Future of Work*. Fieldwork April–May 2024. Luxembourg: Publications Office of the European Union. https://doi.org/10.2767/8591026.

**European Parliament and Council of the European Union. 2024.** Regulation (EU) 2024/1689 of 13 June 2024 laying down harmonised rules on artificial intelligence (Artificial Intelligence Act). *Official Journal of the European Union*, L series, July 12, 2024.

**Eurostat. 2025.** “Individuals – Use of Generative AI Tools [isoc_ai_iaiu].” Eurostat Database. Accessed October 2026. https://ec.europa.eu/eurostat/databrowser/view/isoc_ai_iaiu/default/table.

**Eurostat. n.d.** “Trust in Others by Sex, Age and Educational Attainment [ilc_pw03].” Eurostat Database. Accessed October 2026. https://ec.europa.eu/eurostat/databrowser/view/ilc_pw03/default/table.

**Gillespie, Nicole, Steve Lockey, Alexandria Macdade, Tabi Ward, and Gerard Hassed. 2025.** *Trust, Attitudes and Use of Artificial Intelligence: A Global Study 2025*. Melbourne: University of Melbourne and KPMG. https://doi.org/10.26188/28822919.

**MKOR. 2026.** *AI Adoption Report Romania 2026*. Bucharest: MKOR. Online survey, May–August 2026 (N = 1,000).

**PwC. 2026.** *Danskernes holdning til kunstig intelligens* [Danes’ attitudes to artificial intelligence]. Survey by Epinion, July 2026. Copenhagen: PwC.

**Sandberg, Therese Waalen, Sofie Høllsberg Filstrup, Claes Holger de Vreese, and Arjen van Dalen. 2026.** *Danskerne bruger mere AI, men er skeptiske over for teknologiens udbredelse* *i samfundet: Danskerne, digitalisering og kunstig intelligens 2022–2025*. Odense: Digital Democracy Centre, University of Southern Denmark. https://doi.org/10.21996/f50n-6r25.

**Simon, Felix, Rasmus Kleis Nielsen, and Richard Fletcher. 2025.** *Generative AI and News Report 2025: How People Think About AI's Role in Journalism and Society*. Oxford: Reuters Institute for the Study of Journalism, University of Oxford.